\documentstyle[prc,aps]{revtex}
\begin{document}
\draft
\title{ Nuclear matter properties in relativistic mean field model with 
{\boldmath $\sigma$-$\omega$} coupling}
\author{K.C. Chung$^1$, C.S. Wang$^{1,2}$, A.J. Santiago$^1$, and 
J.W. Zhang$^2$}

\address{\it (1) Instituto de F\'\i sica, Universidade do Estado do Rio 
de Janeiro,\\
Rio de Janeiro-RJ 20559-900, Brazil \\
(2) Department of Technical Physics, Peking University,
Beijing 100871, China}
\date{\today}
\maketitle
\begin{abstract}
The possibility of extending the linear $\sigma$-$\omega$ model by 
introducing a $\sigma$-$\omega$ coupling phenomenologically is explored. 
It is shown that, in contrast to the usual Walecka model, not only the 
effective nucleon mass $M^*$ but also the effective $\sigma$ meson mass 
$m_\sigma^*$ and the effective $\omega$ meson mass $m_\omega^*$ are 
nucleon density dependent. When the model parameters are fitted to the 
nuclear saturation point (the nuclear radius constant $r_0=1.14$fm and 
volume energy $a_1=16.0$MeV) as well as to the effective nucleon mass 
$M^*=0.85M$, the model yields $m_\sigma^*=1.09m_\sigma$ and 
$m_\omega^*=0.90m_\omega$ at the saturation point, and the nuclear 
incompressibility $K_0=501$MeV. The lowest value of $K_0$ given by this 
model by adjusting the model parameters is around $227$MeV.

\end{abstract}


As the starting point for the microscopic relativistic description of 
nuclear many-body system, within the framework of quantum hadrodynamics, 
the well-studied linear $\sigma$-$\omega$ model has been proved to be 
able to describe the saturation and other properties of nuclear 
matter\cite{Serot97}. However, this model yields the nuclear 
incompressibility $K_0$ around $550$MeV which is unacceptably high, and 
also the effective nucleon mass $M^*$ around $0.54M$ which seems 
uncomfortably low. In order to remedy this situation, many works 
have been done on the extension or generalization of this model. Further 
inclusion of nonlinear self-interaction of $\sigma$ mesons has been proved 
to be successful in reducing the nuclear incompressibility $K_0$ 
significantly, in spite of some controversy about the stability of the 
system specified by the fitted parameters\cite{Serot97}\cite{Reinhard89}. 
On the other hand, the density-dependent N-$\sigma$ interaction given by 
the Zimanyi-Moszkowski model\cite{Zimanyi} is shown to be able to give 
reasonable nuclear incompressibility $K_0$ and effective nucleon mass 
$M^*$ as well. Since the Zimanyi-Moszkowski model contains a coupling 
between $\sigma$ and $\omega$ mesons, it is interesting to investigate 
what is the direct effect of the coupling between $\sigma$ and $\omega$ 
mesons, when detached from the density-dependence of the N-$\sigma$ 
coupling. The purpose of this letter is to explore the possibility of 
extending the linear $\sigma$-$\omega$ model by introducing a 
$\sigma$-$\omega$ coupling phenomenologically, and to investigate whether 
this coupling alone is able to reduce the nuclear incompressibility $K_0$ 
and in the same time to increase the effective nucleon mass $M^*$.

Let us start with the following Lagrangian density: 
$${\cal L}=\overline\psi[\gamma_\mu(i\partial^\mu
-g_\omega\omega^\mu)-(M-g_\sigma\phi)]\psi$$
$$+\frac 12\big[(\partial_\mu-\eta g_\omega'\omega_\mu)\phi
(\partial^\mu+\eta g_\omega'\omega^\mu)\phi-m_\sigma^2\phi^2\big]$$
\begin{equation}\label{Lso}-\frac 14F_{\mu\nu}F^{\mu\nu}
+\frac 12m_\omega^2\omega_\mu\omega^\mu,\,\,\,\,
\eta=i\,{\mbox{\rm and}}\,1,
\end{equation}
where $F^{\mu\nu}=\partial^\mu\omega^\nu-\partial^\nu\omega^\mu,$ 
$\psi$, $\phi$ and $\omega$ are the nucleon, $\sigma$ and  $\omega$ 
meson fields with masses $M$, $m_\sigma$ and $m_\omega$, respectively, 
while $g_\sigma$, $g_\omega$ are the respective coupling constants, 
and $g_\omega'$ is the $\sigma$-$\omega$ coupling constant. $\eta=i$ 
gives a $\sigma$-$\omega$ coupling similar to that introduced by usual 
covariant derivatives, while $\eta=1$ gives a gauge-like virtual 
coupling similar to the imaginary coupling\cite{Ai88}. Mathematically, 
it is equivalent to add a term $\eta^2g_\omega'^2\omega_\mu\omega^\mu$ 
to $m_\sigma^2$, which is the same as the $\eta_2$ term introduced in 
the nuclear effective field theory\cite{Furnstahl}. The present model 
reduces to the original Walecka model when the $\sigma$-$\omega$ 
coupling constant is zero, $g_\omega'=0$.

For static nuclear matter, the field equations derived from this 
Lagrangian are reduced to the following equations in the mean field 
approximation:
\begin{equation}\label{Eqpsi}(i\gamma^\mu\partial_\mu
-g_\omega\gamma^0\omega_0-M^*)\psi=0,\end{equation}
\begin{equation}\label{Eqphi}m_\sigma^{*2}\phi=g_\sigma\rho_s,
\end{equation}
\begin{equation}\label{Eqomega}m_\omega^{*2}\omega_0=g_\omega\rho_N,
\end{equation}
where the effective nucleon mass $M^*$, effective $\sigma$ meson mass 
$m_\sigma^*$ and effective $\omega$ meson mass $m_\omega^*$ are defined 
respectively as
\begin{equation}\label{Mstar}M^*=M-g_\sigma\phi,
\end{equation}
\begin{equation}\label{msstar}
m_\sigma^{*2}=m_\sigma^2+\eta^2g_\omega'^2\omega_0^2,
\end{equation}
\begin{equation}\label{mostar}
m_\omega^{*2}=m_\omega^2-\eta^2g_\omega'^2\phi^2,
\end{equation}
and $\rho_s=\langle\overline\psi\psi\rangle$ is the scalar density, 
$\rho_N=\langle\overline\psi\gamma^0\psi\rangle$ the baryon density. For 
calculating these effective masses, the above three equations can be 
rewritten as the following self-consistent equations:
\begin{equation}\label{xiM}
\xi=\frac{M^*}M=1-\frac\alpha{s}\frac{\rho_s}{\rho_0},
\end{equation}
\begin{equation}\label{xis}s=\xi_\sigma^2=\frac{m_\sigma^{*2}}{m_\sigma^2}
=1+\frac{\alpha_\sigma}{v^2}\frac{\rho_N^2}{\rho_0^2},
\end{equation}
\begin{equation}\label{xio}v=\xi_\omega^2=\frac{m_\omega^{*2}}{m_\omega^2}
=1-\frac{\alpha_\omega}{s^2}\frac{\rho_s^2}{\rho_0^2},
\end{equation}
where $\rho_0$ is the standard nucleon number density, $\alpha$, 
$\alpha_\sigma$ and $\alpha_\omega$ are the dimensionless composite 
parameters defined respectively as
\begin{equation}\label{alphaMso}\alpha=\frac{g_\sigma^2\rho_0}{m_\sigma^2M},
\,\,\,\,
\alpha_\sigma=\frac{\eta^2g_\omega'^2g_\omega^2\rho_0^2}{m_\omega^4m_\sigma^2},
\,\,\,\,
\alpha_\omega=\frac{\eta^2g_\omega'^2g_\sigma^2\rho_0^2}{m_\sigma^4m_\omega^2}.
\end{equation}
It should be noted that $\alpha_\sigma$ and $\alpha_\omega$ are positive 
for $\eta=1$ while negative for $\eta=i$.

The nuclear matter equation of state derived from Lagrangian density 
(\ref{Lso}) can be expressed in terms of the nuclear energy density 
${\cal E}$ as $e={\cal E}/\rho_N-M$, and
\begin{equation}\label{Ekso}{\cal E}={\cal E}_k
+{\cal E}_\sigma+{\cal E}_\omega,\end{equation}
\begin{equation}\label{Ek}
{\cal E}_k=\frac{M^4\xi^4}{\pi^2}\sum_{i=p,n}F_1(k_i/\xi M),
\end{equation}
\begin{equation}\label{Esigma}{\cal E}_\sigma=\frac 12(1-\xi)M\rho_s,
\end{equation}
\begin{equation}\label{Eomega}{\cal E}_\omega=\frac 12(1-\xi)yM\rho_s,
\,\,\,\,\,\,y=\frac{(s-1)(2v-1)}{s(1-v)},\end{equation}
where $k_p$ and $k_n$ are the proton and neutron Fermi 
momenta respectively, and the function $F_m(x)$ is defined as
\begin{equation}
F_m(x)=\int_0^xdx\,x^{2m}\sqrt{1+x^2}.
\end{equation}
The baryon density $\rho_N$ and scalar density $\rho_s$ can be expressed as
\begin{equation}\rho_N=\frac 1{3\pi^2}\sum_{i=p,n}k_i^3,
\end{equation}
\begin{equation}\rho_s=\frac{M^3\xi^3}{\pi^2}\sum_{i=p,n}f_1(k_i/\xi M),
\end{equation}
where the function $f_m(x)$ is defined as
\begin{equation}
f_m(x)=\int_0^xdx\frac{x^{2m}}{\sqrt{1+x^2}}.
\end{equation}

Having the equation of state, the pressure $p$ can be derived as
\begin{equation}\label{pressure}p=-{\cal E}
+\rho_N\frac{\partial{\cal E}}{\partial\rho_N}=\frac 13({\cal E}_k
-{\cal E}_\sigma-M\rho_s)+{\cal E}_\omega-\frac{2(1-s)}s{\cal E}_\sigma.
\end{equation}
Instead of Fermi momenta $k_p$ and $k_n$, we will use the nucleon density 
$\rho_N=\rho_n+\rho_p$ and relative neutron excess 
$\delta=(\rho_n-\rho_p)/\rho_N$ as independent variables for describing 
the nuclear matter\cite{Myers69}. At the standard state $\rho_N=\rho_0$, 
$\delta=0$, the pressure should be zero, 
\begin{equation}\label{p0}p\,(\rho_0, 0)=0.\end{equation}
In addition, the depth of the equation of state is related to the nuclear 
volume energy $a_1$ as
\begin{equation}\label{e0}e(\rho_0,0)=-a_1.\end{equation}
The standard nucleon number density $\rho_0$ defined by Eq.(\ref{p0}) is 
related to nuclear radius constant $r_0$ and Fermi momentum $k_F$ of 
standard nuclear matter as $\rho_0=3/4\pi r_0^3=2k_F^3/3\pi^2$.

The generalized nuclear incompressibility $K(\rho_N,\delta)$ 
can be defined\cite{Myers98} and derived as
\begin{equation}\label{K}K\equiv 9\frac{\partial p}{\partial\rho_N}
=\frac{3(p+{\cal E})+12{\cal E}_\omega}{\rho_N}-\Big[3M\rho_s
+\frac{12(3-2s)}s{\cal E}_\sigma\Big]
\frac 1{\rho_s}\frac{\partial\rho_s}{\partial\rho_N}
+\frac{12(3-s)}s\frac{{\cal E}_\sigma}s
\frac{\partial s}{\partial\rho_N}
-\frac{12(1-v)}{1-2v}\frac{{\cal E}_\omega}v
\frac{\partial v}{\partial\rho_N},
\end{equation}
\begin{equation}\rho_N\frac{\partial s}{\partial\rho_N}
=-2(1-s)\Big[1-\frac{2(1-v)}{(1-\xi)v}
\rho_N\frac{\partial\xi}{\partial\rho_N}\Big],
\end{equation}
\begin{equation}\rho_N\frac{\partial v}{\partial\rho_N}
=\frac{2(1-v)}{1-\xi}\rho_N\frac{\partial\xi}{\partial\rho_N},
\end{equation}
\begin{equation}\rho_N\frac{\partial\rho_s}{\partial\rho_N}
=Q+3(\rho_s-Q)\frac{\rho_N}\xi\frac{\partial\xi}{\partial\rho_N},
\end{equation}
\begin{equation}\rho_N\frac{\partial\xi}{\partial\rho_N}
=-\frac 1s\Big[2(1-\xi)(1-s)+\frac{\alpha Q}{\rho_0}\Big]
\Big[1-\frac{4(1-s)(1-v)}{sv}
+\frac{3(1-\xi)(\rho_s-Q)}{\xi\rho_s}\Big]^{-1},
\end{equation}
\begin{equation}Q=\frac{M^3\xi^3}{3\pi^2}
\sum_{i=p.n}\frac{k_i}{\xi M}f_1'(k_i/\xi M).
\end{equation}
The usual nuclear incompressibility $K_0$ can be obtained from 
$K(\rho_N,\delta)$ as  
\begin{equation}\label{K0}K_0=K(\rho_0, 0)=
9\Big(\rho_N^2\frac{\partial^2e}{\partial\rho_N^2}\Big)_0,
\end{equation}
the subscript $0$ stands for the standard state $\rho_N=\rho_0$ and 
$\delta=0$. In addition, the following formula for symmetry energy $J$ can 
be derived:
\begin{equation}\label{J}
J\equiv\frac 12\frac{\partial^2e}{\partial\delta^2}\Big|_0
=\frac 16\frac{k_F^2}{\sqrt{k_F^2+M^2\xi_0^2}}.
\end{equation}

Around the standard state $\rho_N=\rho_0, \delta=0$, the nuclear equation 
of state and thus the nuclear matter properties are specified essentially 
by the standard density $\rho_0$, volume energy $a_1$, symmetry energy 
$J$, incompressibility $K_0$, density symmetry $L$ and symmetry 
incompressibility $K_s$ \cite{Myers69}, where $L$ and $K_s$ are, 
respectively, 
\begin{equation}\label{L}L\equiv\frac 32\Big(\rho_N
\frac{\partial^3e}{\partial\rho_N\partial\delta^2}\Big)_0
=\frac 3{2\rho_0}\frac{\partial^2p}{\partial\delta^2}\Big|_0,
\end{equation}
\begin{equation}\label{Ks}K_s\equiv\frac 92\Big(\rho_N^2
\frac{\partial^4e}{\partial\rho_N^2\partial\delta^2}\Big)_0
=-6L+\frac 12\frac{\partial^2K}{\partial\delta^2}\Big|_0.
\end{equation}

For the description of nuclear matter, there are three independent 
parameters $\alpha$, $\alpha_\sigma$ and $\alpha_\omega$ in the present 
model. The saturation point of nuclear matter ($\rho_0$,$a_1$) defined 
by Eqs.(\ref{p0}) and (\ref{e0}) can be used as input to determine 
$\alpha_\sigma$ and $\alpha_\omega$ as function of $\alpha$. The way is 
as follows. At the standard state ($\rho_N=\rho_0$,\,$\delta=0$), these 
two equations can be solved for $s$ and $y$,
\begin{equation}\label{s}
s=\frac{3(1-\xi)M\rho_s}{3\rho_0(M-a_1)-2{\cal E}_k-\xi M\rho_s},
\end{equation}
\begin{equation}\label{y}
y=\frac{2\rho_0(M-a_1)-2{\cal E}_k-(1-\xi)M\rho_s}{(1-\xi)M\rho_s},
\end{equation}
and thus from Eq.(\ref{Eomega})
\begin{equation}\label{v}v=\frac{sy+s-1}{sy+2s-2}.
\end{equation}
For given $\rho_0$ and $a_1$, $s$ and $v$ can be calculated by 
Eqs.(\ref{s})-(\ref{v}) for a chosen $\xi$. Then $\alpha$, 
$\alpha_\sigma$ and $\alpha_\omega$ can be calculated by 
Eqs.(\ref{xiM})-(\ref{xio}) for a chosen $\xi$. In our calculation, the 
nuclear radius constant $r_0=1.14$fm which corresponds to the standard 
nucleon number density $\rho_0=0.161$fm$^{-3}$, the nuclear volume 
energy $a_1=16.0$MeV, nucleon mass $M=938.9$MeV and constant 
$\hbar c=197.327053$MeV$\cdot$fm are used.

Fig.1 plots $\alpha_\sigma$ and $\alpha_\omega$ as function of $\alpha$. 
At the point $\alpha=\alpha_W=0.4908$ (indicated by an arrow), we have 
$\alpha_\sigma=\alpha_\omega=0$, and the present model reduces to Walecka 
model. For $\alpha>\alpha_W$, $\alpha_\sigma$ and $\alpha_\omega$ are 
negative, which corresponds to $\eta=i$. For $\alpha<\alpha_W$, 
$\alpha_\sigma$ and $\alpha_\omega$ are positive, which corresponds to 
$\eta=1$. There is a lower limit $\alpha_{min}=0.1037$, below this point 
there is no physical solution, as the effective $\omega$ meson mass 
becomes imaginary, $\xi_\omega^2<0$.
Fig.2 shows the dimensionless effective nucleon mass $\xi$, the 
dimensionless effective $\sigma$ meson mass $\xi_\sigma$ and the 
dimensionless effective $\omega$ meson mass $\xi_\omega$ as function of 
$\alpha$. At the point $\alpha_W$, $\xi=0.5437$ and 
$\xi_\sigma=\xi_\omega=1$. From $\alpha_W$ to the right, $\xi$ decreases 
as $\alpha$ increases, and we have $\xi_\sigma<1$ and $\xi_\omega>1$. 
From $\alpha_W$ to the left, $\xi$ increases as $\alpha$ decreases, and 
we have $\xi_\sigma>1$ and $\xi_\omega<1$. 
Fig.3 gives $K_0$, $J$, $L$, and $K_s$ as function of $\alpha$. At 
$\alpha_W$, $K_0=553$MeV, $J=20.2$MeV, $L=70.6$MeV, and $K_s=88$MeV. 
As $\alpha$ decreases, $K_0$ decreases at first and then increases slowly, 
passing through a very flat plateau, and finally decreases rapidly to a 
lower limit $227$MeV. $J$ and $L$ as well as $K_s$ decrease slowly as 
$\alpha$ decreases. It is worthwhile to note that $K_s$ is negative in 
the low $\alpha$ side, in opposition to what is obtained in the usual 
$\sigma$-$\omega$ model. Experimentally, $K_s$ obtained from the 
isoscalar giant monopole resonance energy is between $-566\pm 1350$ to 
$34\pm 159$MeV\cite{Shlomo}. On the other hand, $K_0$ increases to very 
high values rapidly as $\alpha$ increases. In this aspect, the case of 
$\eta=i$ is not acceptable.

The dimensionless effective nucleon mass $\xi$ at the saturation point 
can be used as the third input to fix the parameter $\alpha$. For 
example, if $\xi=0.85$ at the saturation point, we have $\alpha=0.1822$. 
Correspondingly we have $\alpha_\sigma=0.1134$, $\alpha_\omega=0.2938$, 
$\xi_\sigma=1.085$, $\xi_\omega=0.895$, $K_0=501$MeV, $J=13.8$MeV, 
$L=30.1$MeV, and $K_s=-38$MeV, where the values of $\xi_\sigma$ and 
$\xi_\omega$ are those at the saturation point.
Instead of $\xi$, the nuclear incompressibility $K_0$ could be also used 
to fix $\alpha$. If $K_0=250$MeV is chosen, it yields $\alpha=0.1050$, 
$\alpha_\sigma=0.00783$, $\alpha_\omega=2.642$, $\xi=0.939$, 
$\xi_\sigma=1.295$, $\xi_\omega=0.328$, and $K_s=-31.2$MeV, where the 
values of $\xi$, $\xi_\sigma$ and $\xi_\omega$ are those at the saturation 
point.

The symmetry energy $J$ can be increased by including the $\rho$ meson 
field in the present model. In this case, the additional term in the 
Lagrangian density is
\begin{equation}\label{Lrho}{\cal L}_\rho=
-\frac 14{\bf B}_{\mu\nu}\mbox{\boldmath $\cdot$} 
{\bf B}^{\mu\nu}+\frac 12m_\rho^2{\bf b}_\mu\mbox{\boldmath $\cdot$}{\bf b}^\mu
-g_\rho\overline\psi\gamma_\mu\mbox{\boldmath $\tau\cdot$}{\bf b}^\mu\psi,
\end{equation}
where ${\bf B}^{\mu\nu}=\partial^\mu{\bf b}^\nu-\partial^\nu{\bf b}^\mu$, 
${\bf b}^\mu$ is the $\rho$ meson field with mass $m_\rho$ and coupling 
constant $g_\rho$, and {\boldmath $\tau$} are the isospin matrices. The 
resulting nucleon field equation, in the mean field approximation for 
static nuclear matter, has an additional term $-g_\rho\tau_3\gamma^0b_0$:
\begin{equation}\label{Eqpsi1}(i\gamma^\mu\partial_\mu
-g_\omega\gamma^0\omega_0-g_\rho\tau_3\gamma^0b_0-M^*)\psi=0.\end{equation}
The equation for the $\rho$ meson field is
\begin{equation}\label{Eqrho}b_0=-\frac{g_\rho\rho_N\delta}{m_\rho^2},
\end{equation}
while Eqs.(\ref{Eqphi}) and (\ref{Eqomega}) are the same. The additional 
terms to the nuclear energy density, pressure, generalized incompressibility, 
symmetry and density symmetry energies are respectively
\begin{equation}{\cal E}_\rho=\frac 12\alpha_\rho 
M\rho_N^2\delta^2/\rho_0,\,\,\,\,\,\,\,p_\rho={\cal E}_\rho,\,\,\,\,\,\,\,
K_\rho=\frac{18{\cal E}_\rho}{\rho_N},\,\,\,\,\,\,\,
J_\rho=\frac 12\alpha_\rho M,\,\,\,\,\,\,\,L_\rho=3J_\rho,
\end{equation}
where
\begin{equation}\alpha_\rho=\frac{g_\rho^2\rho_0}{m_\rho^2M}
\end{equation}
is the $\rho$ meson dimensionless composite parameter. It is easy to 
prove that there is no contribution to $K_s$ from the $\rho$ meson field.

It can be seen that, as ${\cal E}_\rho=0$ at the standard state where 
$\delta=0$, the inclusion of $\rho$ meson produces no change in nuclear 
energy density ${\cal E}$, pressure $p$ and generalized incompressibility 
$K$ at the standard state. Therefore, the parameters $\alpha$, 
$\alpha_\sigma$ and $\alpha_\omega$ have no change, and thus the 
dimensionless effective masses $\xi$, $\xi_\sigma$ and $\xi_\omega$ are 
the same even the $\rho$ meson field is included. The parameter 
$\alpha_\rho$ can be fixed by using the symmetry energy $J$ as the 
fourth input, if parameters $\alpha$, $\alpha_\sigma$ and $\alpha_\omega$ 
have been fixed.

In conclusion, the possibility of extending the linear $\sigma$-$\omega$ 
model by introducing a $\sigma$-$\omega$ coupling phenomenologically is 
explored. It is shown that, in contrast to the usual Walecka model, not 
only the effective nucleon mass $M^*$ but also the effective $\sigma$ 
meson mass $m_\sigma^*$ and the effective $\omega$ meson mass 
$m_\omega^*$ are nucleon density dependent. When the model parameters 
are fitted to the nuclear radius constant $r_0=1.14$fm and volume energy 
$a_1=16.0$MeV as well as to the effective nucleon mass $M^*=0.85M$, the 
model yields $m_\sigma^*=1.09m_\sigma$ and $m_\omega^*=0.90m_\omega$ 
at the same nuclear saturation point, and the nuclear incompressibility 
$K_0=501$MeV. This incompressibility seems too high. On the other hand, 
if the model parameters are fitted to $K_0=250$MeV, in addition to 
$r_0=1.14$fm and $a_1=16.0$MeV, it yields $M^*=0.938M$, 
$m_\sigma^*=1.295m_\sigma$ and $m_\omega^*=0.328m_\omega$ at the nuclear 
saturation point. This effective $\omega$ meson mass seems too low. In 
addition, the lower limit of incompressibility $K_0=227$MeV is not low 
enough. Therefore, even this model is able to reduce the nuclear 
incompressibility and also to increase the effective nucleon mass 
simultaneously, there is still not enough degrees of freedom to adjust 
the parameters in order to give more reasonable result. In this aspect, 
additional physics should be included for improving this model.

\begin{figure}
\caption{$\alpha_\sigma$ and $\alpha_\omega$ as function of $\alpha$. 
The point $\alpha_W=0.4908$ indicated by an arrow corresponds to the 
usual Walecka model where $\alpha_\sigma=\alpha_\omega=0$.}
\label{Figure1}
\end{figure}

\begin{figure}
\caption{The dimensionless effective nucleon mass $\xi$, the 
dimensionless effective $\sigma$ meson mass $\xi_\sigma$ and the 
dimensionless effective $\omega$ meson mass $\xi_\omega$ as function of 
$\alpha$. At the point $\alpha_W=0.4908$ indicated by an arrow, 
$\xi=0.5437$ and $\xi_\sigma=\xi_\omega=1$.}
\label{Figure2}
\end{figure}

\begin{figure}
\caption{$K_0$, $J$, $L$, and $K_s$ as function of $\alpha$. At the point 
$\alpha_W=0.4908$ indicated by an arrow, $K_0=553$MeV, $J=20.2$MeV, 
$L=70.6$MeV, and $K_s=88$MeV. At the lower limit $\alpha_{min}=0.1037$, 
$K_0=227$MeV.}
\label{Figure3}
\end{figure}

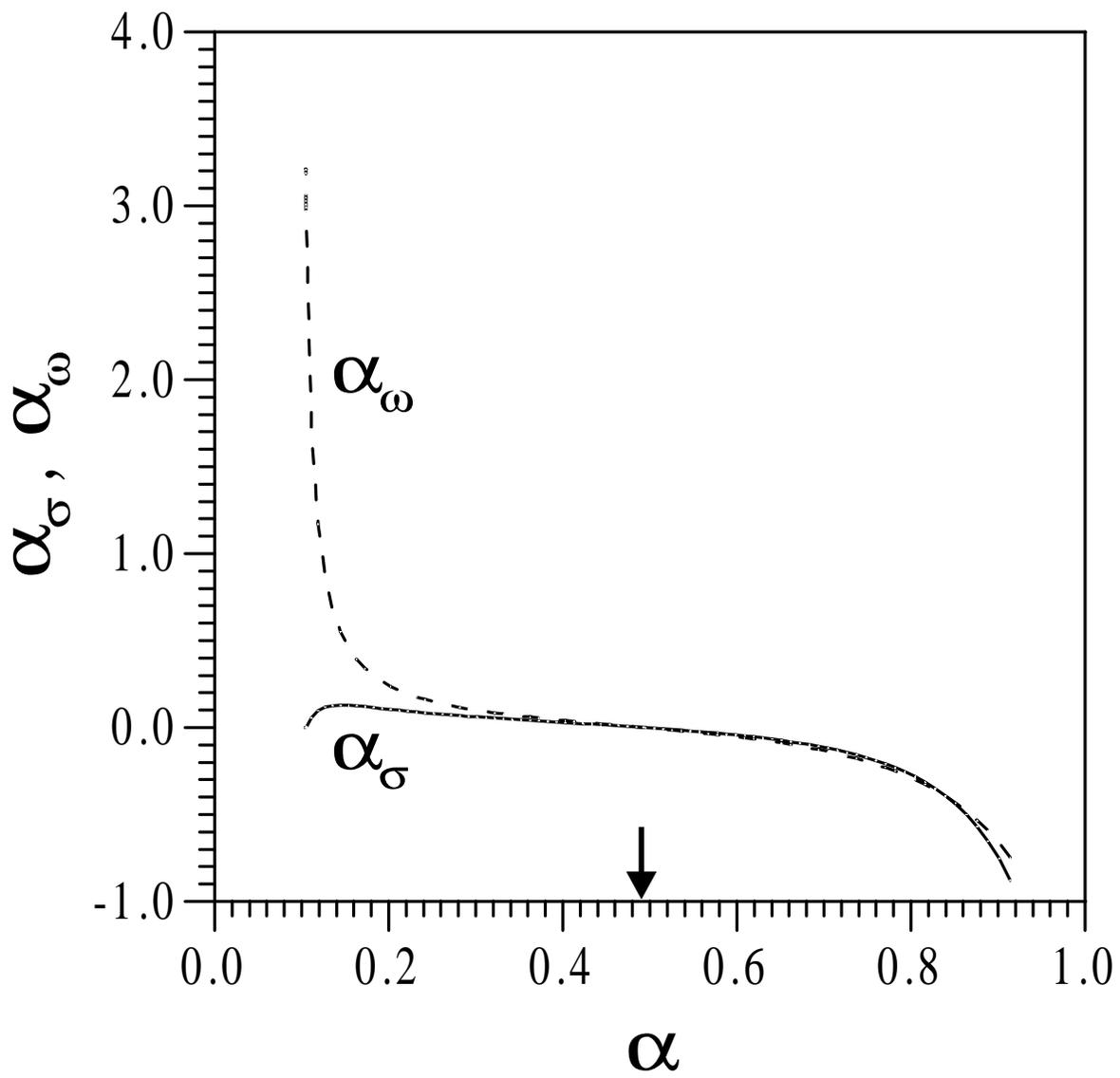
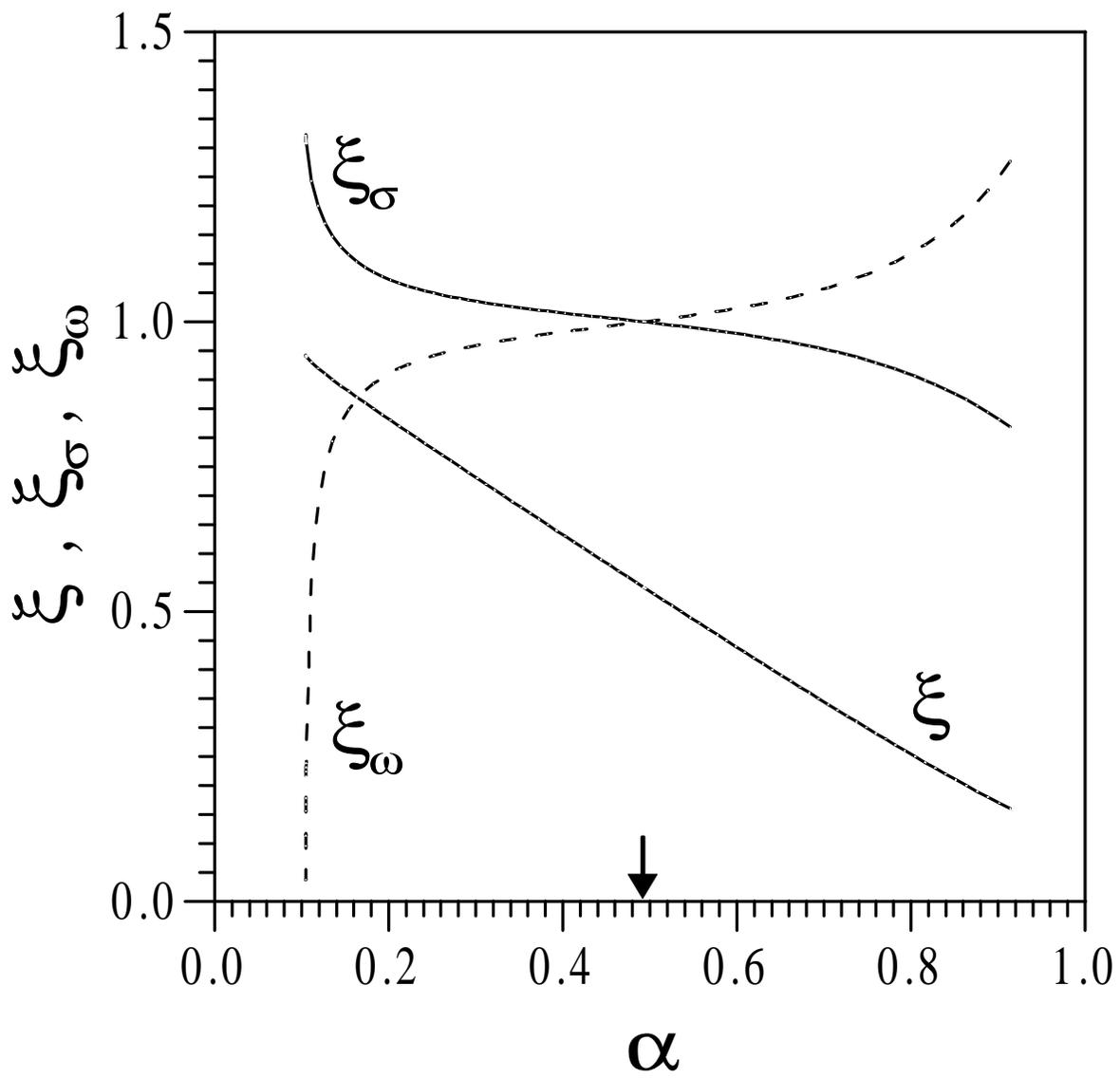
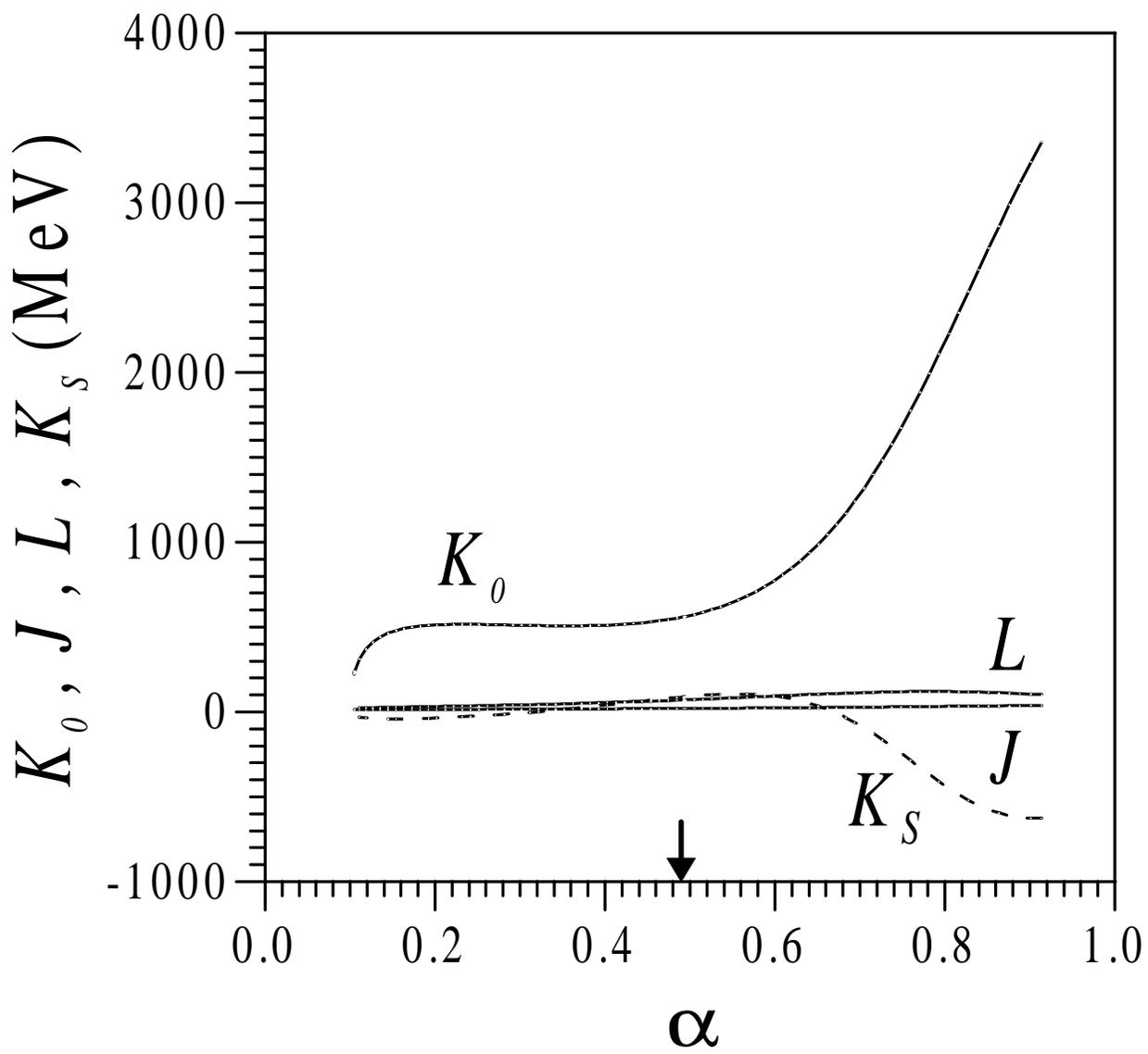
\end{document}